\documentclass[]{aa}

\usepackage[varg]{txfonts}
\usepackage[]{graphicx}

\usepackage{natbib}
\bibpunct{(}{)}{;}{a}{}{,} % to follow the A&A style

\begin{document}

\title{Ejection of close-in super-Earths around low-mass stars in the giant impact stage}

\author{
	Yuji Matsumoto \inst{\ref{inst:1}}
	\and Pin-Gao Gu \inst{\ref{inst:1}}
	\and Eiichiro Kokubo \inst{\ref{inst:3}}
	\and Shoichi Oshino \inst{\ref{inst:4}}
	\and Masashi Omiya \inst{\ref{inst:5}}
	}

\institute{
	Institute of Astronomy and Astrophysics, Academia Sinica, Taipei 10617, Taiwan\\
	\email{ymatsumoto@asiaa.sinica.edu.tw}\label{inst:1}
	%\and
	%Institute of Astronomy and Astrophysics, Academia Sinica, Taipei 10617, Taiwan
	%\label{inst:2}
	\and
	National Astronomical Observatory of Japan, 2-21-1, Osawa, Mitaka, 181-8588 Tokyo, Japan
	\label{inst:3}
	\and
	Institute for Cosmic Ray Research, University of Tokyo, Hida, Gifu 506-1205, Japan
	\label{inst:4}
	\and
	Astrobiology Center, NINS, 2-21-1, Osawa, Mitaka, 181-8588 Tokyo, Japan
	\label{inst:5}
	}

\date{Received dd mm yyyy / Accepted dd mm yyyy}

\abstract
{%Context
Earth-sized planets were observed in close-in orbits around M dwarfs.
While more and more planets are expected to be uncovered around M dwarfs, theories of their formation and dynamical evolution are still in their infancy. 
}
{%Aim
We investigate the giant impact growth of protoplanets, which includes strong scattering around low-mass stars.
The aim is to clarify whether strong scattering around low-mass stars affects the orbital and mass distributions of the planets.
}
{%Methods
We perform $N$-body simulation of protoplanets by systematically surveying the parameter space of the stellar mass and surface density of protoplanets.
}
{%Results
We find that protoplanets are often ejected after twice or three times close-scattering around late M dwarfs.
The ejection sets the upper limit of the largest planet mass.
Adopting the surface density scaling linearly with the stellar mass, we find that as the stellar mass decreases less massive planets are formed in orbits with higher eccentricities and inclinations.
Under this scaling, we also find that a few close-in protoplanets are generally ejected.
}
{%Conclusion
The ejection of protoplanets plays an important role in the mass distribution of super-Earths around late M dwarfs.
The mass relation of observed close-in super-Earths and their central star mass is well reproduced by ejection.
}


\titlerunning{Ejection of close-in super-Earths}
\authorrunning{Y. Matsumoto et al.}

\maketitle

\section{Introduction} \label{sec:intro}

\begin{figure}[hbt]
	\includegraphics[width=\columnwidth]{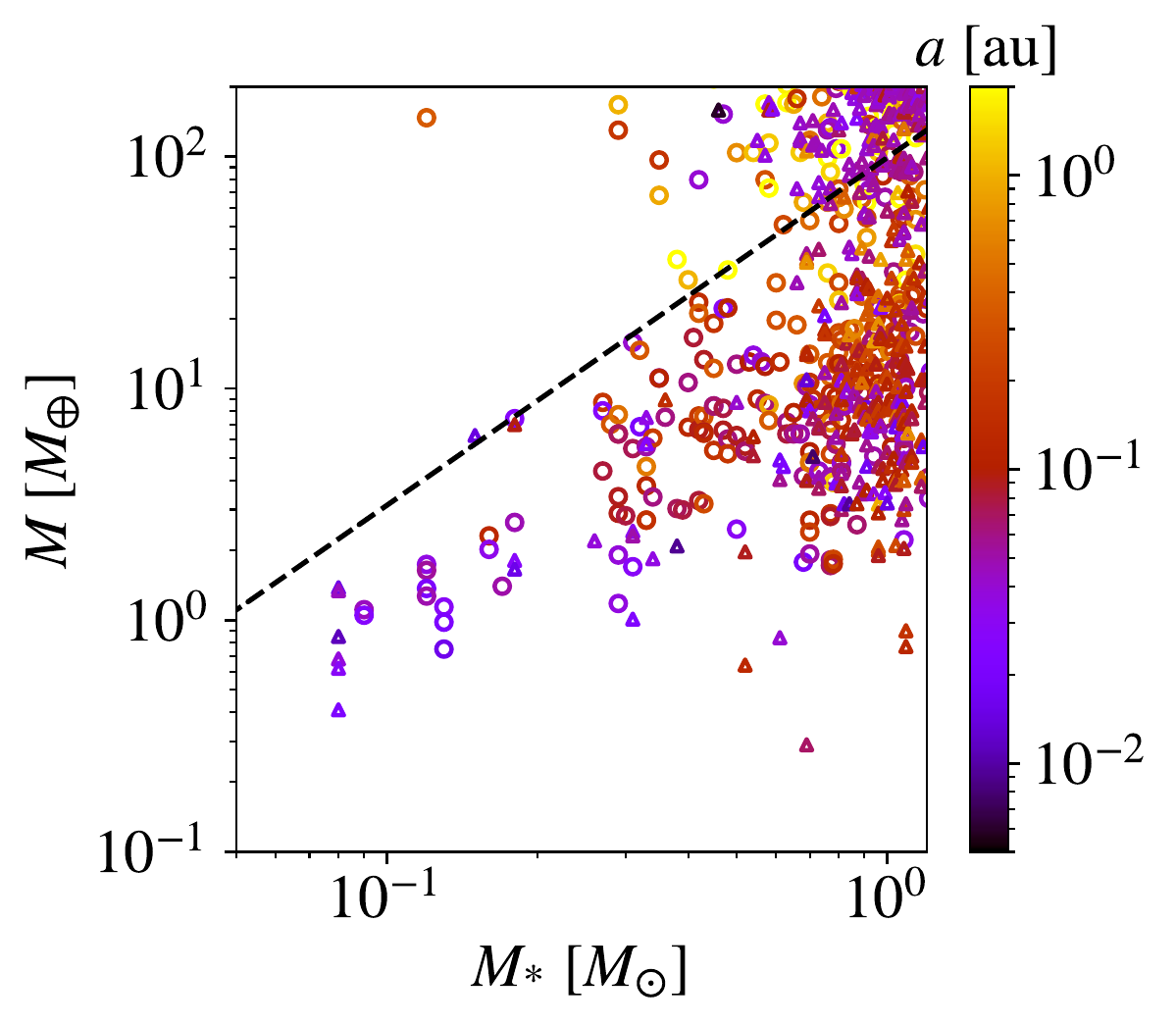}
	\caption{
		Distribution of observed planets on the stellar mass in the solar mass unit ($M_*$ [$M_{\odot}$]) and planetary mass in the Earth mass unit ($M$ [$M_{\oplus}$]) plane.
		The data was extracted from the NASA exoplanet archive (https://exoplanetarchive.ipac.caltech.edu/) as of April 2020.
		Point colors represent semimajor axis of each planet.
		Planets shown in circles are detected by the radial velocity (RV) method and those in triangles are transit or transit timing methods.
		The dashed line is the ejection mass at 0.1~au given by Equation (\ref{eq:M_ej}).
		}
	\label{Fig:Ms_M_obs}
\end{figure}

Recent observations have reported the presence of planets around M dwarfs with mass of $\sim0.1M_{\odot}$, where $M_{\odot}$ is the Solar mass \citep{Anglada-Escude+2016, Gillon+2017,Zechmeister+2019}.
These planets are located within 0.1~au, and their masses are about Earth mass.
In addition to these planets, almost dozens of planets are found around stars with $\lesssim 0.3M_{\odot}$ \citep[e.g.,][]{Hirano+2016a,Angelo+2017,Bonfils+2018}.
Since planet occurrence rates tend to be high around low-mass stars \citep{Dressing&Charbonneau2013,Dressing&Charbonneau2015, Hardegree-Ullman+2019}, on-going and coming observational missions are expected to find more and more planets around M dwarfs.
Figure \ref{Fig:Ms_M_obs} shows the mass relation of observed planets and their central stars.
This figure clearly shows that low-mass stars hold small mass planets and there are no $\sim 10$ Earth mass planets around $\sim0.1M_{\odot}$ stars, while such mass planets are abundant around $1M_{\odot}$ stars.
In this paper, we focus on the mass relation between the stars and these planets, which have $\lesssim 10 M_{\oplus}$ and are located at $\lesssim 1$~au, where $M_{\oplus}$ is the Earth mass.
We call these planets as close-in super-Earths.
It is worth noting that this mass gap of close-in super-Earths exists around $\lesssim0.5M_{\odot}$ stars and there are no planets reported by {\it Kepler} mission around such stars since these planets are orbiting around $>0.5M_{\odot}$ stars \citep{Johnson+2017,Weiss+2018b,Weiss+2018}.

The formation of close-in super-Earths is composed of various processes \citep[e.g.,][]{Raymond+2014, Raymond&Morbidelli2020}.
Their formation can be divided into two stages: formation of protoplanets and their giant impact growth. 
Protoplanets grow by accretion of planetesimals \citep[e.g.,][]{Wetherill&Stewart1989,Kokubo&Ida1996,Kokubo&Ida1998, Kokubo&Ida2000} and/or pebbles \citep[e.g.,][]{Ormel&Klahr2010, Lambrechts&Johansen2012, Johansen&Lambrechts2017}.
These protoplanets originally grow either in the close-in orbit \citep[e.g.,][]{Ogihara+2015a, Ogihara&Hori2020} or in the outer region \citep[e.g.,][]{Kley&Nelson2012, Rein2012}. 
Due to the planet-disk tidal interaction \citep[e.g.,][]{Ward1986,Tanaka+2002,Ida+2020b}, they are trapped into resonant chains around the disk inner edge.
When the number of protoplanets in the resonant chain is small, they do not cause giant impacts, and planets are formed in resonant chain orbits. 
When their number is large, they cause orbital instabilities and giant impacts occur \citep[e.g.,][]{Ogihara&Ida2009,Matsumoto+2012, Matsumoto&Ogihara2020}.
We note that most planets are considered to have evolved according to the later scenario since they are not in resonant chains \citep{Fabrycky+2014, Winn&Fabrycky2015}.
The distribution of protoplanets affects the distribution of planets since it determines whether giant impacts occur. 
Recent studies showed that the distributions of protoplanets and subsequently formed planets are affected by the main accretion source (planetesimals or pebbles) and gas removal processes \citep[e.g.,][]{Ogihara+2018, Izidoro+2017,Izidoro+2019, Ogihara&Hori2020}.

Most of these theoretical studies considered the formation of close-in super-Earths around $1M_{\odot}$ stars, and the distribution of protoplanets around M dwarfs is not well known.
Recently, \citet{Liu_B+2019, Liu_B+2020} showed that small planets are formed around low-mass stars when they grow via pebble-driven core accretion. 
These studies focused on the formation stage of protoplanets.
They considered the growth of a single protoplanet, and the distribution of protoplanets in a system is not clear.

The giant impact growth is also mainly studied around $1M_{\odot}$ stars \citep[e.g.,][]{Kokubo+2006, Hansen&Murray2012, Hansen&Murray2013, Matsumoto&Kokubo2017}.
The dynamical process in the giant impact stage around M dwarfs is not well known since there are no studies considering the systematical survey on the parameter space composed of the stellar mass and protoplanet mass, while there are some studies considering the dependence of the stellar mass on the giant impact growth as follows.
\cite{Lissauer2007} performed $N$-body simulations of planetesimals and protoplanets around the habitable zone of $(1/3)M_{\odot}$.
The planetary accretion around the habitable zone of M dwarfs was also investigated by \cite{Raymond+2007}.
They considered the stellar mass as a parameter ranging from $0.2M_{\odot}$ to $1M_{\odot}$.
\cite{Ciesla+2015} also performed simulations in the same stellar mass range.
\cite{Montgomery&Laughlin2009} considered the growth of planetary embryos around $0.12M_{\odot}$ stars. 
\cite{Ogihara&Ida2009} performed $N$-body simulations of planetesimals, considering orbital migration around $0.2M_{\odot}$.
\cite{Moriarty&Ballard2016} examined the planetary accretion around $1M_{\odot}$ and $0.5M_{\odot}$ to reproduce the distribution of planets that exhibits the so-called Kepler dichotomy found by the {\it Kepler} mission \citep{Johansen+2012}.
 
In this paper, we focus on the in-situ formation of close-in super-Earths via giant impacts around low-mass stars to consider the planet mass distribution around low-mass stars (Fig. \ref{Fig:Ms_M_obs}).
We systematically change the stellar mass between $1M_{\odot}$ and $10^{-5/4}M_{\odot}\simeq 0.056M_{\odot}$.
The lowest stellar mass in our parameter range is in the mass range of brown dwarfs \citep{Reid&Hawley2005}.
We initially put isolation mass protoplanets in a gas-free disk \citep{Kokubo&Ida2000,Kokubo&Ida2002}.
The corresponding disk surface densities are $10$ -- $100$~g~cm$^{-2}$ at 1~au.
While these densities are more massive than the surface densities suggested by observations \citep[e.g.,][]{Williams&Cieza2011, Andrews+2013}, we use these surface densities to reproduce $\sim 1M_{\oplus}$ to $\sim 10M_{\oplus}$ planets \citep[see][for more detailed comparisons of the observed dust mass and observed planet mass]{Manara+2018}.
As the stellar mass decreases, scattering between protoplanets becomes relatively strong for a given mass of protoplanets.
Our results show that such strong scattering causes protoplanet ejections from M dwarfs.
These ejection events give a possible explanation for the planet-stellar mass relation around low-mass stars: why there are no $\sim 10$ Earth mass planets around $\sim0.1M_{\odot}$ stars.

This paper is planned as follows.
Our numerical model is described in Sect. \ref{sect:model}.
We show how the stellar mass and surface density affect the final orbital and mass distributions of planets in Sect. \ref{sect:results}.
Section \ref{sect:summary} presents the conclusion.

\section{Numerical model}\label{sect:model}

\subsection{Initial Conditions}\label{sect:ic}

We consider the giant impact growth of protoplanets around a star whose mass is $M_*$ without disk gas.
The stellar mass ($M_*$) ranges from $1M_{\odot}$ to $10^{-5/4}M_{\odot}$.
This parameter range of the stellar mass includes the mass range of M dwarfs \citep{Reid&Hawley2005, Kaltenegger&Traub2009} and the ultra-cool stars where planets have been observed \citep{Gillon+2017,Zechmeister+2019}.

The mass distribution of protoplanets that are precursors of close-in super-Earths is still under debate.
In this paper, we put protoplanets, which have isolation masses \citep{Kokubo&Ida2000, Kokubo&Ida2002}.
We note that we do not imply that protoplanets in close-in orbits are formed according to the oligarchic model.
We adopt the isolation mass model since this model gives us a relation between protoplanet masses and surface density.
The surface density of protoplanets is given by the power-law disk model, $\Sigma=\Sigma_1 (a/\mbox{1~au})^{-3/2}$ with the power index taken from the minimum-mass solar nebular model \citep{Hayashi1981}.
This power-law index is slightly larger than those derived from observed close-in super-Earths by \citet{Chiang&Laughlin2013} ($-1.6$) and \citet{Dai_F+2020} ($-1.75$).
The protoplanets are initially located at $a=(0.05\mbox{~au},1\mbox{~au})$, where $a$ is the semimajor axis of a protoplanet.
We place the innermost protoplanet at $0.05 \mbox{~au}+ 0.5b_{\rm init}$.
The semimajor axes of the other planets are given by $a_{i+1}=a_i+b_{\rm init}$ within 1~au.
The initial orbital separation is a parameter and its fiducial value is $b_{\rm init}=10r_{\rm H}$, where $r_{\rm H}$ is the Hill radius of a protoplanet ($r_{\rm H}=( 2M/3M_* )^{1/3}a$).
The initial masses of protoplanets are given by \citep{Kokubo&Ida2000, Kokubo&Ida2002}
\begin{eqnarray}
	M &\simeq& 2\pi a b_{\rm init} \Sigma \nonumber \\
	&\simeq& 0.16 M_{\oplus}\left( \frac{a}{1\mbox{~au}} \right)^{3/4}
	\left( \frac{b_{\rm init}}{10 r_{\rm H}} \right)^{3/2}
	\left( \frac{\Sigma_1}{10 \mbox{~g~cm}^{-2}} \right)^{3/2}
	%\nonumber\\&&\times 	
	\left( \frac{M_*}{M_{\odot}} \right)^{-1/2}
	.
	\label{eq:Miso}
\end{eqnarray}
The total mass of the initial protoplanets ($M_{\rm tot,init}$) is almost equal to 
\begin{eqnarray}
	M_{\rm tot,init} %&\simeq& 2\pi\int^{a_{\rm out}}_{a_{\rm in}} \Sigma ada 
	\simeq 3.7M_{\oplus} \left( \frac{\Sigma_1}{10 \mbox{~g~cm}^{-2}} \right) ,
	\label{eq:Mtot}
\end{eqnarray}
which is obtained by an integration of the surface density between 0.05~au and 1~au.
The surface density at 1~au ($\Sigma_1$) is a parameter in our simulations.
In this study, we focus on the mass distribution of planets between $\sim M_{\oplus}$ and $\sim 10M_{\oplus}$ around low-mass stars.
As \cite{Matsumoto&Kokubo2017} showed, it is expected that several similar mass planets are formed in close-in orbits.
Accordingly, we take $\Sigma_1=$ 10, 30, and 100~g~cm$^{-2}$.

\begin{table}
	\caption{Summary of models.}\label{table:model}
	\begin{tabular}{llll}
		\hline\hline
		Model	&	$M_*/M_{\odot}$	&	$\Sigma_1$~[g~cm$^{-2}$]	&	$b_{\rm init}/r_{\rm H}$\\
		\hline
		Ms0$\Sigma$100b10	&	$1.0$		&	100	&	10\\
		Ms0$\Sigma$30b10	&	$1.0$		&	30	&	10\\
		Ms0$\Sigma$10b10	&	$1.0$		&	10	&	10\\
		Ms1$\Sigma$100b10	&	$10^{-1/4}$	&	100	&	10\\
		Ms1$\Sigma$30b10	&	$10^{-1/4}$	&	30	&	10\\
		Ms1$\Sigma$10b10	&	$10^{-1/4}$	&	10	&	10\\
		Ms2$\Sigma$100b10	&	$10^{-1/2}$	&	100	&	10\\
		Ms2$\Sigma$30b10	&	$10^{-1/2}$	&	30	&	10\\
		Ms2$\Sigma$10b10	&	$10^{-1/2}$	&	10	&	10\\
		Ms3$\Sigma$100b10	&	$10^{-3/4}$	&	100	&	10\\
		Ms3$\Sigma$30b10	&	$10^{-3/4}$	&	30	&	10\\
		Ms3$\Sigma$10b10	&	$10^{-3/4}$	&	10	&	10\\
		Ms4$\Sigma$100b10	&	$0.1$		&	100	&	10\\
		Ms4$\Sigma$30b10	&	$0.1$		&	30	&	10\\
		Ms4$\Sigma$10b10	&	$0.1$		&	10	&	10\\
		Ms5$\Sigma$100b10	&	$10^{-5/4}$	&	100	&	10\\
		Ms5$\Sigma$100b8	&	$10^{-5/4}$	&	100	&	8\\
		Ms5$\Sigma$100b6	&	$10^{-5/4}$	&	100	&	6\\
		Ms5$\Sigma$30b10	&	$10^{-5/4}$	&	30	&	10\\
		Ms5$\Sigma$10b10	&	$10^{-5/4}$	&	10	&	10\\
		\hline
	\end{tabular}
\end{table}

Each model in our simulations is named after $M_*$, $\Sigma_1$, and $b$.
For instance, Ms0$\Sigma$100b10 is the model for $M_*=10^0M_{\odot}$, $\Sigma_1=100$~g~cm$^{-2}$, and $b_{\rm init}=10r_{\rm H}$.
All models in our simulations are summarized in Table \ref{table:model}.
We calculate 20 runs in each model by randomly changing the initial angles of protoplanets.

As the stellar mass decreases, the mass of the initial protoplanet increases (see Equation (\ref{eq:Miso})), while the number of initial protoplanets ($n_{\rm init}$) decreases since the Hill radius increases ($r_{\rm H}\propto M^{1/3}M_*^{-1/3}$) and $n_{\rm init}$ is roughly proportional to $b_{\rm init}^{-1}$.
The number of initial protoplanets is the smallest, $n_{\rm init}=6$, in the Ms5$\Sigma$100b10 model, while it is the largest, $n_{\rm init}=67$, in the Ms0$\Sigma$10b10 model.
To evaluate the effects of $n_{\rm init}$ and initial $M$ on the results, we also perform simulations with $b_{\rm init}=8r_{\rm H}$ and $6r_{\rm H}$ for $M_*=10^{-5/4}M_{\odot}$ and $\Sigma_1=100$~g~cm$^{-2}$ (models Ms5$\Sigma$100b8 and Ms5$\Sigma$100b6 in Table \ref{table:model}).
%e and i
The initial eccentricities ($e$) and inclinations ($i$ [rad]) are given by the Rayleigh distribution with dispersions $\langle e^2 \rangle^{1/2}=2\langle i^2 \rangle^{1/2} = r_{\rm H}/a$, which are proportional to $M_*^{-1/3}$. 
It would be worth noting that \cite{Matsumoto&Kokubo2017} showed that initial eccentricities of protoplanets do not affect results and initial inclinations affect results only when they change in orders of magnitudes.

\subsection{Orbital Integration}\label{sect:integration}

The orbital motions of protoplanets are calculated by the direct integration of their equations of motion.
We use the fourth-order Hermite scheme \citep{Makino&Aarseth1992,Kokubo&Makino2004} with the hierarchical timestep \citep{Makino1991}.
The maximum timesteps are almost equal to $0.01T_{\rm K}$ of the innermost planet, where $T_{\rm K}$ is the Keplerian time.
This is about 1~hr for a planet at 0.05~au around a $1M_{\odot}$ star and 0.2~hr for that at 0.05~au around a $10^{-5/4}M_{\odot}$ star.
Such short timesteps enable us to precisely resolve collisions of protoplanets, which are important for collisional damping of eccentricities and inclinations \citep{Matsumoto+2015, Matsumoto&Kokubo2017}.
Adopting this timescale, the error in the total energy in $T_{\rm K}$ of the innermost planet is $\mathcal{O}(10^{-12})$ when we put one protoplanet.

We follow the evolution for at least $10^7$ yr.
After $10^7$~yr, we estimate the orbital crossing time of the system using the minimum orbital crossing time derived by \cite{Ida&Lin2010} (see also Appendix \ref{sect:delta--ecc}).
We continue the integration until the crossing time becomes longer than $10^7$~yr or the orbital elements become almost constant in the last $10^7$~yr.
In some calculations, we continue simulations over $10^8$~yr and confirm that they do not lead to orbital instabilities.
This is because the accretion timescale in the giant impact stage is $\sim 10^8T_{\rm K}$ -- $10^9T_{\rm K}$ \citep[][]{Agnor+1999, Kokubo+2006, Hansen&Murray2012,Moriarty&Ballard2016}.
We assume perfect accretion; namely, when protoplanets collide, they always merge.
Protoplanets have the same bulk density, $\rho=3~\mbox{g~cm}^{-3}$.

\section{Results}\label{sect:results}

\subsection{Typical evolution}\label{sect:typical}

\begin{figure*}[ht]
	\resizebox{\hsize}{!}{
		\includegraphics{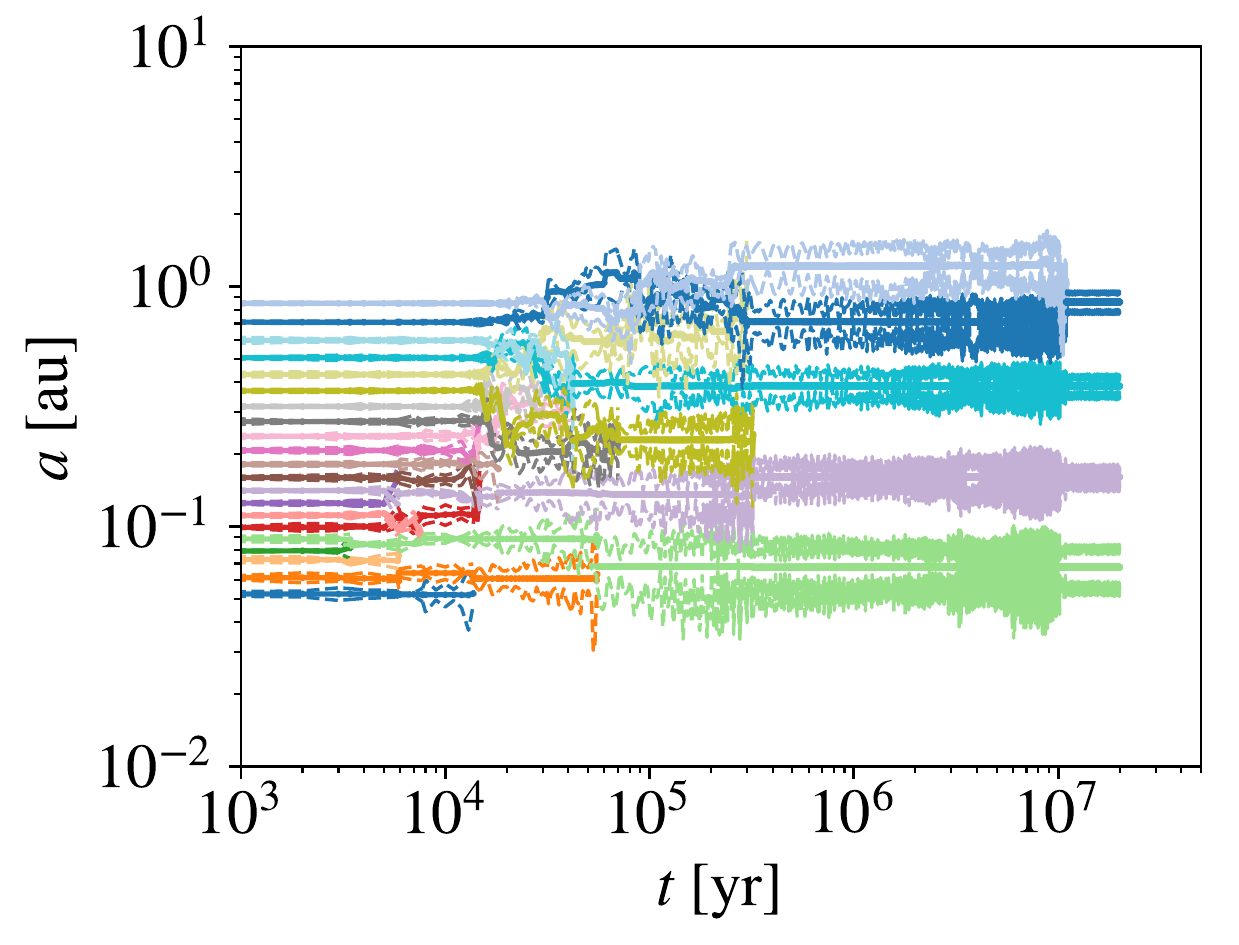}
		\includegraphics{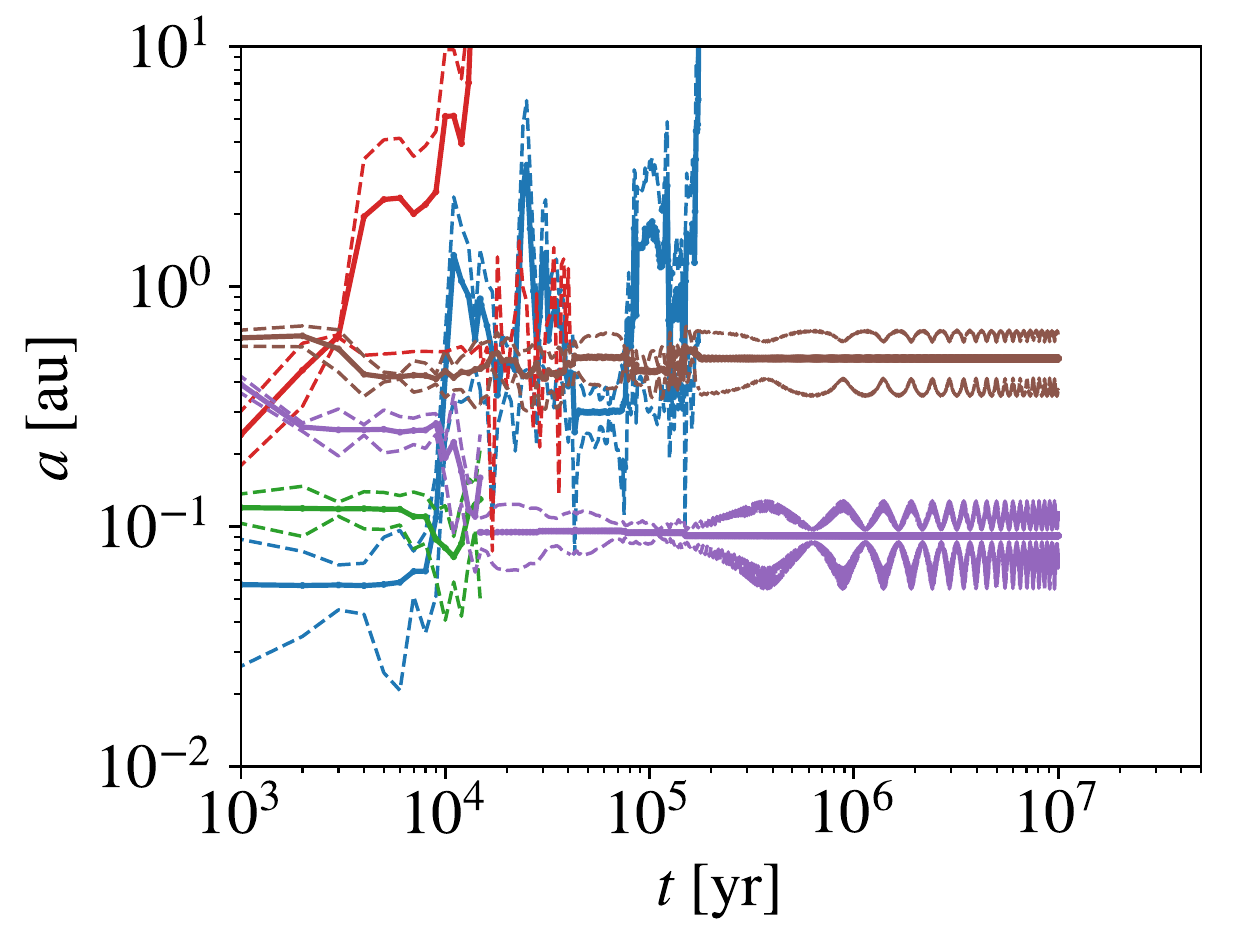}
	}
	\caption{
		Typical time evolution is shown for models Ms0$\Sigma$100b10 (left, $M_*=1M_{\odot}$) and Ms5$\Sigma$100b10 (right, $M_*=10^{-5/4}M_{\odot}$) is shown.
		The solid and dashed curves in the same color show the semimajor axis and pericenter and apocenter distances of each body.
		When two protoplanets collide, the merged body has the same color as that of the massive protoplanet.
	}
	\label{Fig:typical_oe}
\end{figure*}

Although the disk properties are the same, the orbital evolution of protoplanets becomes significantly different when we change the stellar mass.
Figure \ref{Fig:typical_oe} shows the time evolution of semimajor axes and pericenter and apocenter distances in models Ms0$\Sigma$100b10 (left panel, $M_*=1M_{\odot}$) and Ms5$\Sigma$100b10 (right one, $M_*=10^{-5/4}M_{\odot}$), respectively.
In these models, the total masses of the initial protoplanets are about $37M_{\oplus}$.
In the Ms0$\Sigma$100b10 model, orbital crossing occurs from smaller orbital radii due to the shorter orbital periods.
Most of the collisions occur soon after orbital crossing, which results in the tournament-like structure of the time evolution of semimajor axes \citep{Matsumoto&Kokubo2017}.
In this case, four planets are finally formed.
The mass of the innermost planet is $5.1M_{\oplus}$ and those of the other three planets are between $10M_{\oplus}$ and $12M_{\oplus}$.
The eccentricities ($e$) and inclinations ($i$) of the planets are 0.15, 0.097, 0.076, 0.10 and 0.091~rad, 0.0514~rad, 0.12~rad, and 0.088~rad from the innermost, respectively.
The estimated crossing timescale of the final planets is longer than $10^{10}$ yr.
Except for the innermost scattered planet, planets have similar mass and small eccentricities ($\leq 0.1$).
The final eccentricities of planets are given by the close-scattering and collisional damping when inclinations of initial protoplanets are small \citep{Matsumoto+2015,Matsumoto&Kokubo2017}.
The eccentricity raised in a close-scattering event is given by the fraction of the escape velocity to the Kepler velocity, 
\begin{eqnarray}
	e_{\rm esc} 
	%&=& \sqrt{\frac{2(M_k+M_l)}{M_*} \frac{a}{R_k+R_l} } \nonumber\\
	&\simeq& 0.19
		\left(\frac{M_i+M_j}{10M_{\oplus}} \right)^{1/3} 
		\left( \frac{a}{0.1~\mbox{au}} \right)^{1/2} 
		\left( \frac{\rho}{3 \mbox{~g~cm}^{-3}} \right)^{1/6}
		\left( \frac{M_*}{M_{\odot} }\right)^{-1/2}
		,
		\nonumber\\
		%&&\times 
	\label{eq:e_esc}
\end{eqnarray}
which is often called escape eccentricity
\citep[][]{Safronov1972, Kokubo&Ida2002, Leinhardt&Richardson2005, Pfyffer+2015,Matsumoto+2015, Matsumoto&Kokubo2017}.
The eccentricity of the scattered innermost planet is $\sim e_{\rm esc}$, and those of the other planets are $\sim 0.5e_{\rm esc}$ due to the collisional $e$-damping.

Such a tournament-like structure is not seen in the Ms5$\Sigma$100b10 model.
Scattering between protoplanets becomes more prominent as the stellar mass decreases since their Hill radii become larger.
Protoplanets tend to scatter each other rather than collide.
Strong scattering leads to protoplanet ejections.
In the right panel of Fig. \ref{Fig:typical_oe}, two ejection events occur.
The first ejection occurs at $4.2\times 10^4$~yr.
While the semimajor axis of the first ejected body (the red solid curve) exceeds 10~au at $1.3\times 10^4$~yr, its pericenter (the red dashed curve) stays at $\lesssim 1$~au since its eccentricity is $e>0.9$.
Its eccentricity exceeds unity due to scattering with other protoplanets at $4.2\times 10^4$~yr.
The second ejection happens at $1.8\times 10^5$~yr.
The semimajor axis of the second ejected body (the blue solid curve) changes drastically several times due to scattering.
This body is ejected by multiple scattering with the outermost planet (the brown curve).
Finally, two planets are formed. 
The inner one has $M=19M_{\oplus}$, $e=0.18$, and $i=0.41$~rad.
The outer one has $M=15M_{\oplus}$, $e=0.26$, and $i=0.22$~rad. 
These final eccentricities of planets are less than $0.3e_{\rm esc}$. 

The dynamical evolution of close-in super-Earths around $\sim 0.1M_{\odot}$ stars is similar to that of giant planets around $1M_{\odot}$ stars \citep[e.g.,][]{Rasio&Ford1996,Marzari&Weidenschilling2002,Nagasawa+2008}.
Strong scattering between protoplanets leads to ejection.
Around low-mass stars, gravitational scattering between protoplanets is strong owing to their large initial mass (Equation (\ref{eq:Miso})) and the weak stellar gravity.
Protoplanets around low-mass stars easily gain high eccentricities ($e_{\rm esc}\propto M^{1/3} M_*^{-1/2}$), and some are subsequently ejected.
Although the two ejected bodies (the red and blue ones) acquire high eccentricities $e\sim e_{\rm esc}>0.9$, they are not immediately ejected after their orbits are crossed.
Their pericenters stay around the semimajor axis of the outermost body (the brown one).
These are scattered again, and their eccentricities exceed unity.
Therefore, protoplanets are ejected by multiple close-scattering even when $e_{\rm esc}$ is smaller than unity.
The giant impact growth of protoplanets is regulated by ejection due to multiple scattering events as the stellar mass becomes low.

\subsection{The collision-dominated regime and ejection-dominated regime}
\label{sect:dependence_Ms}

\begin{figure*}[hbt]
	\resizebox{\hsize}{!}{
		\includegraphics{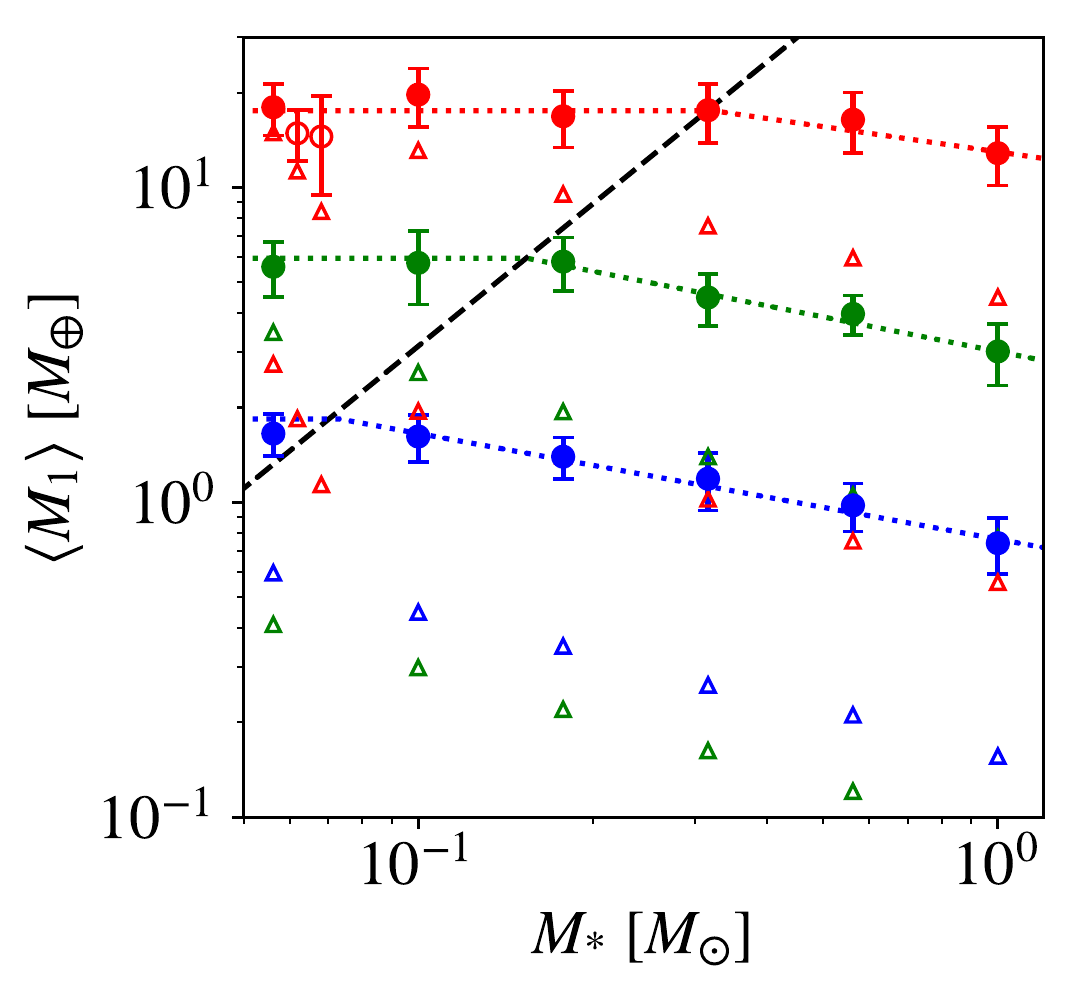}
		\includegraphics{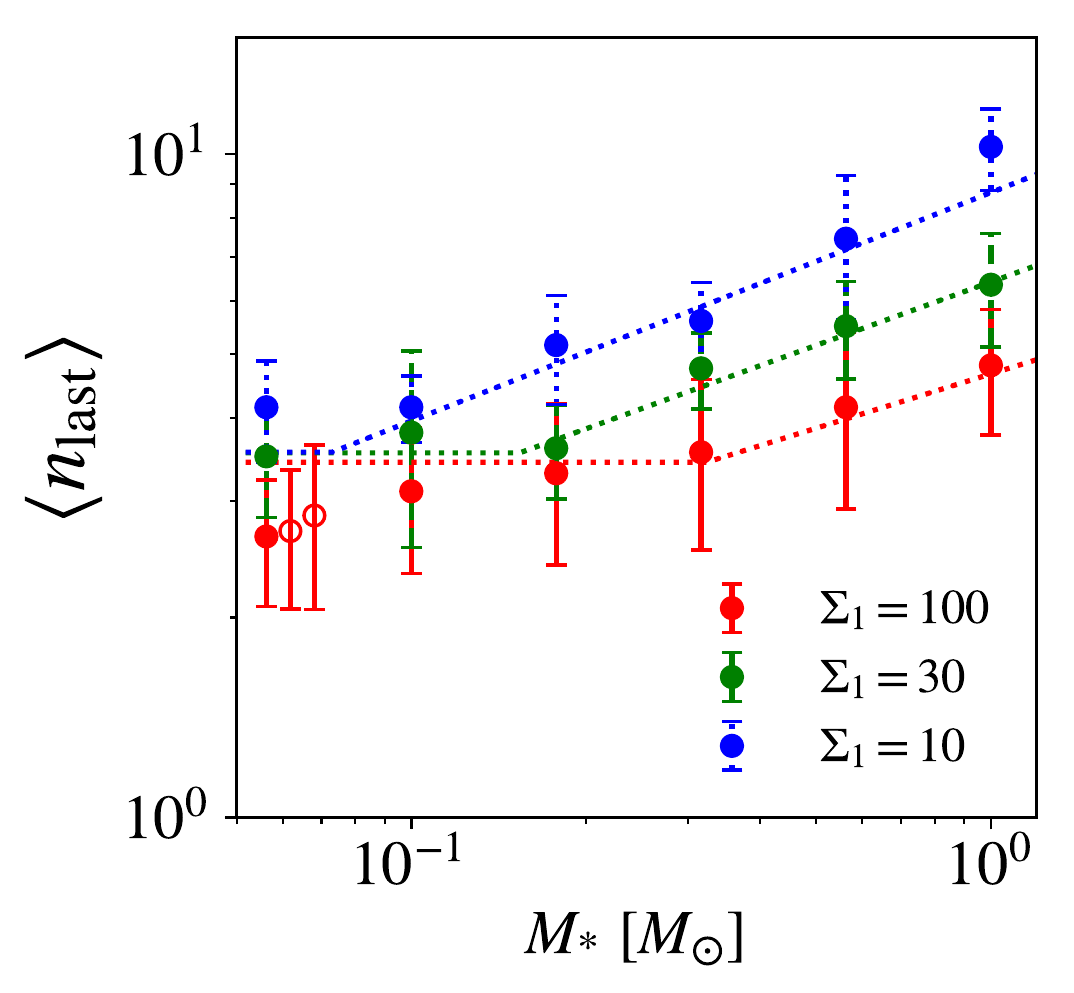}
	}
	\caption{
		Average values of the most massive planets ($\langle M_1\rangle$) and the number of planets ($\langle n_{\rm last}\rangle$) in a system are shown as a function of the stellar mass.
		The length of the error bars is equal to the standard deviation.
		We plot the results of $\Sigma_1=10~\mbox{g~cm}^{-2}$ in blue filled symbols,
		$\Sigma_1=30~\mbox{g~cm}^{-2}$ in green filled ones, and
		$\Sigma_1=100~\mbox{g~cm}^{-2}$ in red filled ones.
		The open circles are the results of models Ms5$\Sigma$100b8 (left) and Ms5$\Sigma$100b6 (right), which are plotted at slightly large $M_*$ to avoid overlapping.
		In the left panel, the dashed line is the ejection mass ($M_{\rm ej}$) given by Equation (\ref{eq:M_ej}).
		The triangles are the mass of the largest and smallest initial protoplanets.
		The dotted lines are the fitting lines.
		In the collision-dominated regime ($\langle M_1\rangle<M_{\rm ej}$), we derive the fitting lines from the mass of the most massive planets and the final number of planets in all simulation results.
		In the ejection-dominated regime ($\langle M_1\rangle>M_{\rm ej}$), these lines are constant and equal to those values at $\langle M_1\rangle=M_{\rm ej}$.
		}
	\label{Fig:Ms_M1_n}
\end{figure*}

\begin{table*}
	\caption{
		Coefficients of $\langle M_1/M_{\oplus} \rangle$ and $\langle n_{\rm last}\rangle$ in Equation (\ref{eq:fit}).
		}\label{table:coefficients}
	\begin{tabular}{lllll}
	\hline\hline
	$y$	&		&	$\Sigma_1=10~\mbox{g~cm}^{-2}$	&	$\Sigma_1=30~\mbox{g~cm}^{-2}$	&	$\Sigma_1=100~\mbox{g~cm}^{-2}$\\
	\hline
	$\langle M_1/M_{\oplus} \rangle$	
			&	$C_{M1}$	&	-0.30$\pm$0.026	&	-0.36$\pm$0.033	&	-0.27$\pm$0.062\\
			&	$C_{M2}$	&	-0.092$\pm$0.017	&	0.48$\pm$0.016	&	1.1$\pm$0.020\\
			&	$C_{M3}$	&	0.26	&	0.78	&	1.2\\
			&	$\log{\langle M_1/M_{\oplus} \rangle} $	&	0.21$\pm$0.067	&	0.74$\pm$0.10	&	1.2$\pm$0.091\\
		\hline
		$n_{\rm last}$	
			&	$C_{n1}$	&	0.31$\pm$0.025	&	0.32$\pm$0.030	&	0.27$\pm$0.069\\
			&	$C_{n2}$	&	0.92$\pm$0.017	&	0.81$\pm$0.014	&	0.67$\pm$0.022\\
			&	$C_{n3}$	&	0.55	&	0.55	&	0.54\\
			&	$ \log{\langle n_{\rm last} \rangle} $	&	0.60$\pm$0.085	&	0.55$\pm$0.11	&	0.46$\pm$0.12\\
	\hline
	\end{tabular}
\end{table*}

As the stellar mass decreases, $e_{\rm esc}$ increases and planetary scattering becomes more effective. 
This promotes planetary growth via giant impacts.
However, when $e_{\rm esc}$ exceeds a certain value, the ejection becomes effective and planets do not grow via giant impacts.
These features can be seen in Fig. \ref{Fig:Ms_M1_n}.
The left and right panels of Fig. \ref{Fig:Ms_M1_n} show the mass of the most massive planets ($\langle M_1\rangle$) and the final number of planets ($\langle n_{\rm last}\rangle$) as the function of the stellar mass.
These panels suggest that there are two regimes.
For the high stellar mass, $\langle M_1\rangle$ increases and $\langle n_{\rm last}\rangle$ decreases as the stellar mass decreases.
We call this regime the collision-dominated regime, which means that the collisional growth is dominant for planetary growth, and protoplanets grow as $e_{\rm esc}$ increases in this regime.
The mass of the most massive planet and the number of final planets become almost constant when the stellar mass becomes less than a certain value since the ejection is dominant.
This is the ejection-dominated regime.
The boundary of these regimes is estimated by the ejection condition, 
\begin{eqnarray}
	M_{\rm ej} &\simeq& 3.0 M_{\oplus}
		\left( \frac{e_{\rm esc}}{0.4}\right)^{3}
		\left( \frac{a}{0.1~\mbox{au}} \right)^{-3/2} 
		\left( \frac{\rho}{3 \mbox{~g~cm}^{-3}} \right)^{-1/2}
		\left( \frac{M_*}{0.1 M_{\odot} }\right)^{3/2}
		.
		\nonumber\\ 
	\label{eq:M_ej}
\end{eqnarray}
We plot Equation (\ref{eq:M_ej}) in the left panel of Fig. \ref{Fig:Ms_M1_n} (the black dashed line).
The boundary is well reproduced by $M_{\rm ej}$ with $e_{\rm esc}=0.4$.
This $e_{\rm esc}$ value implies that protoplanets are ejected by two or three close-scatterings.

In the ejection-dominated regime, the collision growth of protoplanets is suppressed by the strong scattering. 
If the initial protoplanets are smaller than the ejection mass, the growth of protoplanets would be stalled at the ejection mass.
This is a possible explanation for the absence of close-in super-Earths with the mass $>M_{\rm ej}$ around low-mass stars (Fig. \ref{Fig:Ms_M_obs}).

In the following, we investigate the dependencies of $\langle M_1\rangle$ and $\langle n_{\rm last}\rangle$ on the stellar mass and the surface density in detail.
We give the relationship between the planetary mass, stellar mass, and surface density of protoplanets.
We compare our results to \cite{Kokubo+2006} where the dependencies of $\langle M_1\rangle$ and $\langle n_{\rm last}\rangle$ on the surface density are discussed.
In the collision-dominated regime ($\langle M_1\rangle\leq M_{\rm ej}$), $\langle M_1\rangle$ and $\langle n_{\rm last}\rangle$ are expressed as the power-law functions of the stellar mass.
These become constatnt in the ejection-dominated regime ($\langle M_1\rangle>M_{\rm ej}$).
Using $y= M_1/M_{\oplus}$ or $ n_{\rm last}$, their stellar mass dependencies are expressed as
\begin{eqnarray}
	\log{y} &=& 
		\left\{
			\begin{array}{lr}
				C_{y1} \log{\left( M_*/M_{\odot} \right)} + C_{y2} ,&	(\langle M_1\rangle \leq M_{\rm ej}),\\
				C_{y3},&(\langle M_1\rangle > M_{\rm ej}),
			\end{array}
		\right. 
	\label{eq:fit}
\end{eqnarray}
where $C_{y1}$ and $C_{y2}$ are fitting coefficients of each quantity ($\langle M_1/M_{\oplus}\rangle$ and $ n_{\rm last}$) and $C_{y3}$ is derived from Equation (\ref{eq:fit}) at $\langle M_1\rangle=M_{\rm ej}$.
These values are summarized in Table \ref{table:coefficients}.
The power-law indices of the largest mass ($C_{M1}$) are equal to -1/3 within almost $1\sigma$ deviations in all $\Sigma_1$ models.
This can be interpreted as the dependence of the Hill radius on the stellar mass.
The power-law indices of the number of the final planets ($C_{n1}$) are approximately equal to $-C_{M1}$.
This is because of mass conservation.
As planets become massive, the final number of planets decreases.

The intercept $C_{M2}$, which is equal to $\log{\langle M_1/M_{\oplus}\rangle }$ in $1M_{\odot}$ models, is well expressed by $\Sigma_1$ and given by 
\begin{eqnarray}
	C_{M2}=1.2\log{\left( \frac{\Sigma_1 }{ 10~\mbox{g~cm}^{-2} }\right)}-0.093.
	\label{eq:C_M2}
\end{eqnarray}
This surface density dependence is almost identical to that in \cite{Kokubo+2006}, where the surface density dependence of the largest planet mass is $1.1\log{(\Sigma_1/10~\mbox{g~cm}^{-2})}+0.079$.
The most massive planets in our $M_*=M_{\odot}$ models have smaller masses than those in \cite{Kokubo+2006}, which considered the giant impact growth of planets around 1 au.
This is because collisions are more effective rather than scattering and several (4.8--9.3) smaller planets of similar mass form in close-in orbits while a few ($\lesssim 4$) larger planets are formed around 1~au \citep{Matsumoto+2015, Matsumoto&Kokubo2017}.
This can be also interpreted by $e_{\rm esc}$, which is an increasing function of the semimajor axis.
The $e_{\rm esc}$ values in \cite{Kokubo+2006} are about 0.3.
As $e_{\rm esc}$ increases, the mass of the most massive planets increases due to strong scattering.

The number of the final planets in the collision-dominated regime is also well expressed by the power-law function of $\Sigma_1$,
\begin{eqnarray}
	C_{n2}&=&-0.25\log{\left( \frac{\Sigma_1}{10~\mbox{g~cm}^{-2}} \right) }+0.92.
	\label{eq:C_n2}
\end{eqnarray}
The number of the final planets can be determined by the relation between the total mass ($\propto \Sigma_1$) and the most massive planets ($\propto \Sigma_1^{1.2}$, Equation (\ref{eq:C_M2})) based on mass conservation in the collision-dominated regime.

We derive $C_{M3}$ from Equation (\ref{eq:fit}) with $\langle M_1\rangle=M_{\rm ej}$ and $\log{\langle M_1/M_{\oplus} \rangle} $ directly from our results.
These values are almost the same.
In the ejection-dominated regime,
\begin{eqnarray}
	\log{\left\langle \frac{M_1}{M_{\oplus}} \right\rangle } =1.0\log{\left( \frac{\Sigma_1}{10~\mbox{g~cm}^{-2} } \right)}+0.23.
\end{eqnarray}
The dependence of $\log{\langle M_1/M_{\oplus} \rangle}$ on the surface density in this regime becomes slightly weaker than that in the collision-dominated regime.
The mass of the most massive planets in the ejection-dominated regime is given by the condition that $\langle M_1\rangle$ in the collision-dominated regime is equal to $M_{\rm ej}$.
As $\Sigma_1$ decreases, this condition is satisfied around lower mass stars (Equation (\ref{eq:M_ej}), Fig. \ref{Fig:Ms_M1_n}).
This makes the dependence of $\log{\langle M_1/M_{\oplus} \rangle }$ on $\Sigma_1$ slightly weaker than $C_{M2}$.
We also derive $C_{n3}$ and $\log{\langle n_{\rm last} \rangle }$,
\begin{eqnarray}
	\log{ \langle n_{\rm last} \rangle }=-0.14\log{\left( \frac{\Sigma_1}{10~\mbox{g~cm}^{-2} } \right)}+0.60,
\end{eqnarray}
which is a similar dependence to \cite{Kokubo+2006}.

While $C_{n3}$ is within 1$\sigma$ values of $\log{ \langle n_{\rm last} \rangle }$, the number of the final planets in $\Sigma_1=100~\mbox{g~cm}^{-2}$ case is overestimated (see points and fitting line of $\Sigma_1=100\mbox{~g~cm}^{-2}$ models in Fig. \ref{Fig:Ms_M1_n}).
In these models, most of the initial protoplanets have larger masses than the ejection mass.
Ejections frequently occur and the number of planets becomes small.
These features can be confirmed by the results of models Ms5$\Sigma$100b8 and Ms5$\Sigma$100b6, where the initial protoplanets are less massive than those in the Ms5$\Sigma$100b10 model.
The numbers of the final planets are slightly larger than that in the Ms5$\Sigma$100b10 model since ejection becomes less effective.
This suggests that planets grow only via a few collisions and these collisions determine the final planetary mass in the ejection-dominated regime.
This implies that the growth of protoplanets would be stalled at the ejection mass if the masses of the initial protoplanets are smaller than the ejection mass.

We get the dependencies of $M_1$, which are proportional to $M_*^{-1/3} \Sigma_1^{1.2}$ in the collision-dominated regime and $\Sigma_1$ in the ejection-dominated regime. 
These dependencies are different from that of the isolation mass, which is proportional to $M_*^{-1/2}\Sigma_1^{3/2}$. 
Moreover, the ejection mass would set the upper limit of $M_1$ when protoplanets are less massive than $M_{\rm ej}$. 
These reflect the dynamical process in the giant impact growth.

\subsection{Collision velocities}\label{sect:discuss_col}

\begin{figure}[hbt]
	\resizebox{\hsize}{!}{
		\includegraphics{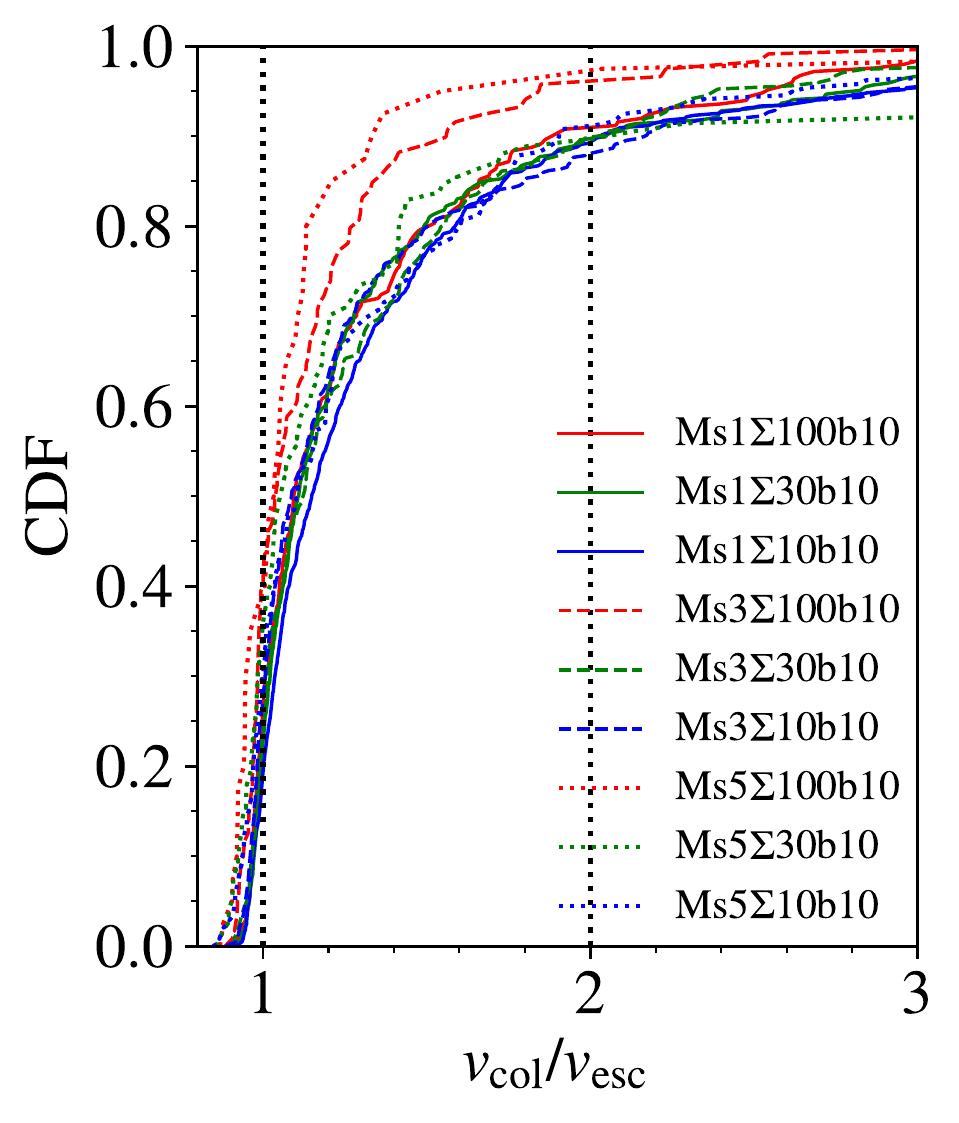}
		\includegraphics{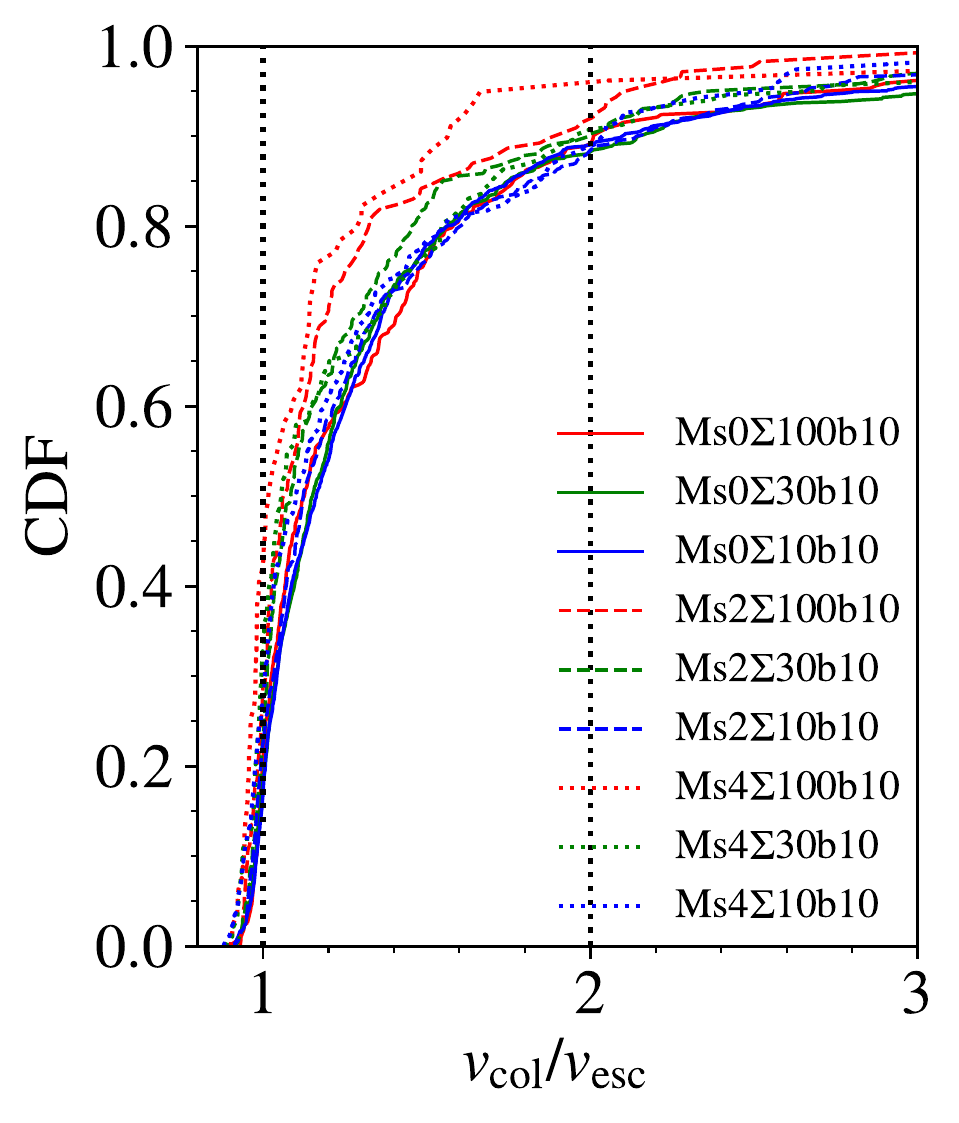}
	}
	\caption{
		Cumulative distribution of the collision velocities normalized by the escape velocity.
		Vertical black dotted lines are $v_{\rm col}=v_{\rm esc}$ and $v_{\rm col}=2v_{\rm esc}$.
		}
	\label{Fig:vcol_vesc_CDF}
\end{figure}

While we assume the perfect accretion, collisional outcomes depend on collision velocities.
The cumulative distributions of the collision velocities are shown in Fig. \ref{Fig:vcol_vesc_CDF}.
In most of the models, these distributions are similar. 
In the Ms0$\Sigma$100b10 and Ms1$\Sigma$100b10 models, the average collision velocities normalized by the escape velocities ($v_{\rm col}/v_{\rm esc}$) are $1.37 \pm 0.60$ and $1.31\pm 0.53$, repspectively.
In $\Sigma_1=100\mbox{~g~cm}^{-2}$ models, the collision velocities becomes lower as the stellar mass decreases.
While high $v_{\rm col}/v_{\rm esc}$ collisions occur when protoplanets experience some close scattering events before collisions, protoplanets are ejected after a few close scattering events in the ejection-dominated regime.
Thus, high velocity collisions occur less frequently in the ejection-dominated regime.
This is why the distribution of the collision velocities fall primarily on the small value around low-mass stars.

Most of the collisions in our simulations do not cause erosion but perfect accretion, partial accretion, graze-and-merge, and hit-and-run \citep{Asphaug2010, Leinhardt&Stewart_ST2012, Stewart&Leinhardt2012}.
The properties of collisions in our simulations are similar to the previous studies that consider $N$-body simulations with realistic collisions around 1 solar mass stars \citep{Kokubo&Genda2010,Chambers2013, Poon+2020}.
These studies found that there are no substantial differences from simulations with only perfect accretion.
This is because the energy dissipation in hit-and-run collisions induces a next collision that leads to accretion \citep[][]{Emsenhuber&Asphaug2019,Emsenhuber&Asphaug2019b}.
Our results do not change significantly if we consider these realistic collisions.

\subsection{Distribution of planets around M dwarfs}\label{sect:distribution}

\subsubsection{Mass relation between massive super-Earths and stars}\label{sect:hd}

We investigate the observable planets using a certain relation between the surface density and the stellar mass.
The dependence of the surface density of protoplanets on the stellar mass ($\Sigma_1\propto M_*^{h_{\rm d} }$) is not known.
We adopt $h_{\rm d}=1$ according to the recent study about the surface density of observed planets \citep[][$h_{\rm d}=1.04$]{Dai_F+2020}. 
This dependence is also adopted in the fiducial case of \cite{Raymond+2007} and slightly shallower than that of the protoplanetary mass formed via pebble accretion \citep[][$h_{\rm d}=4/3$]{Liu_B+2019}.
Under this dependence, we consider two series of models.
Models Ms1$\Sigma$100b10, Ms3$\Sigma$30b10 and Ms5$\Sigma$10b10 correspond to $\Sigma_1\simeq 100\mbox{~g~cm}^{-2}(M_*/10^{-1/4}M_{\odot})$, and models Ms0$\Sigma$100b10, Ms2$\Sigma$30b10 and Ms4$\Sigma$10b10 correspond to $\Sigma_1\simeq 100\mbox{~g~cm}^{-2}(M_*/M_{\odot})$.
We name the former series as the massive series and the latter series as the less-massive series.

These series well reproduce the mass distribution of the observed massive super-Earths (Fig. \ref{Fig:Ms_M_obs}).
In the massive series, the mass of the most massive planets is estimated to be $22M_{\oplus}$ around $1M_{\odot}$ stars and $3.0M_{\oplus}$ around $0.1M_{\odot}$ stars (Sect. \ref{sect:dependence_Ms}, Equation (\ref{eq:M_Ms_h1})).
These masses agree with the typical mass of the observed massive super-Earths \citep[e.g.,][]{Butler+2006,Anglada-Escude+2016, Gillon+2017,Zechmeister+2019}. 
We note that we focus on the distribution of massive super-Earths, which are observable around low-mass stars by on-going and coming observational missions of RV measurements. 
The surface densities in the massive and less-massive series are similar to those estimated by RV planets and larger than those by California-Kepler-Survey \citep{Dai_F+2020}.

\subsubsection{Orbits and masses of planets}\label{sect:orbit_mass}

\begin{figure*}[htb]
	\resizebox{\hsize}{!}{
		\includegraphics{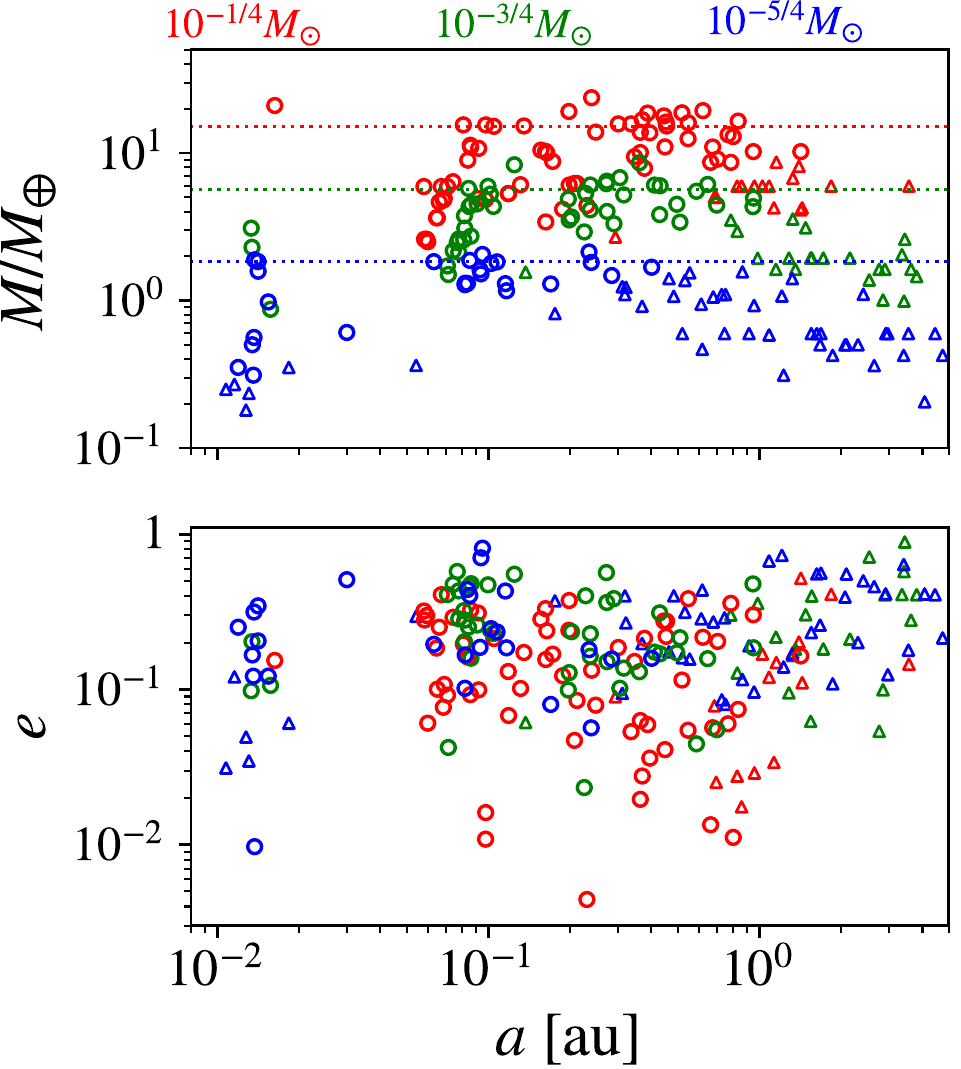}
		\includegraphics{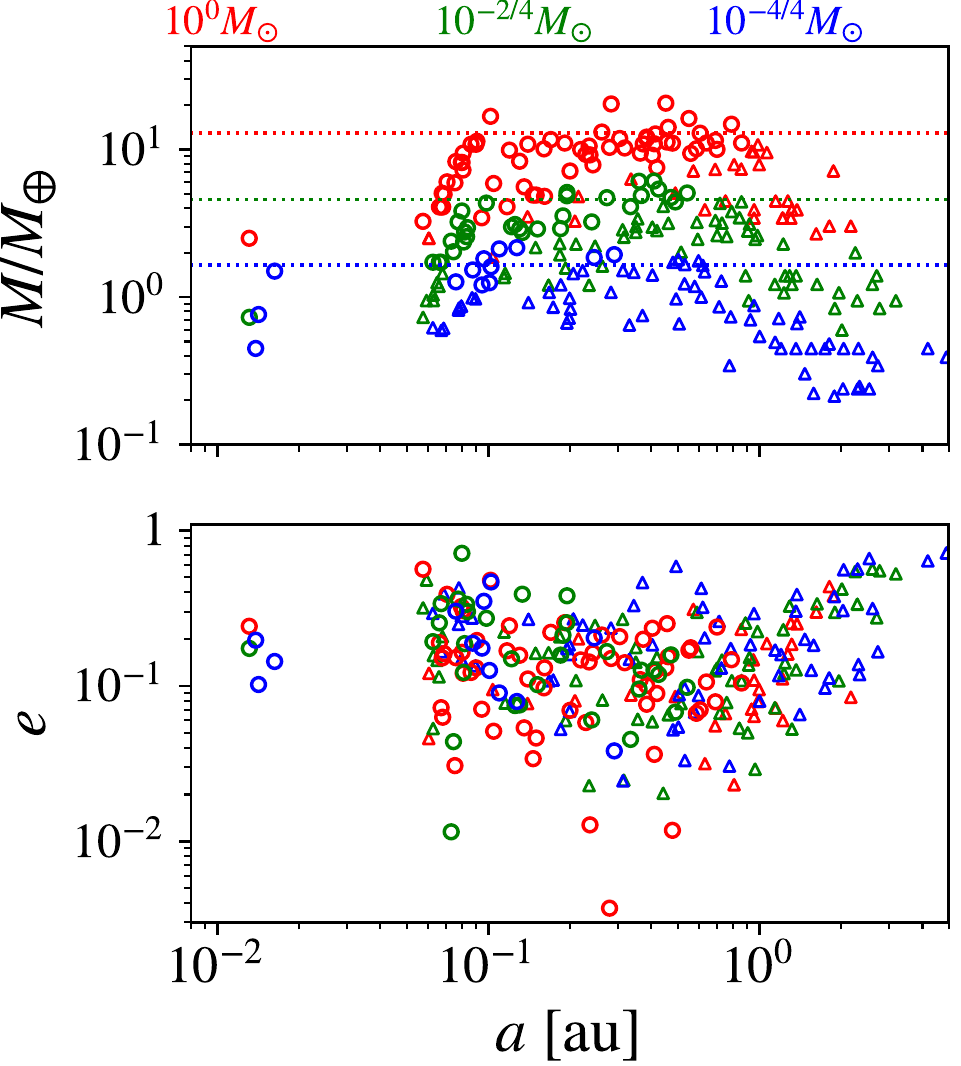}
		}
	\caption{
		Snapshots of the final planets plotted on the $a-M$ (top) and $a-e$ (bottom) planes.
		The planets represented by circles and triangles have larger and smaller radial velocity amplitudes than 1~m~s$^{-1}$, respectively.
		The dotted lines in the top panels are given by Equation (\ref{eq:fit}) for $M_1/M_{\oplus}$.
		In the left panels, we plot the results of the massive series, which include models Ms1$\Sigma$100b10 (red symbols), Ms3$\Sigma$30b10 (green ones), and Ms5$\Sigma$10b10 (blue ones).
		In the right panels, we plot the results of the less-massive series, which are models Ms0$\Sigma$100b10 (red symbols), Ms2$\Sigma$30b10 (green ones), and Ms4$\Sigma$10b10 (blue ones).
		}
	\label{Fig:formed_aMei}
\end{figure*}

Figure \ref{Fig:formed_aMei} shows the distribution of the final planets in each series.
The most massive planets are formed between $\sim 0.1$~au and $\sim1$~au, which correspond to the initial distribution of the protoplanets.
Less massive planets are scattered and more widely distributed.
The eccentricities of scattered planets are higher than the massive ones.
These two groups (the massive planets and scattered planets) exhibit the bimodal distribution of planetary masses.
These features can be seen in both series. 

Several planets are located at $\sim 10^{-2}$~au, while most innermost planets are located at $0.05$ -- $0.1$~au, which is almost the same location of the initial innermost protoplanets.
When a planet is scattered to the inner orbit by a strong scattering associated with an ejection event, a planet is pushed into the inner orbit than the orbit of the innermost protoplanet.
This dynamical process was already found in the previous studies, which considered the dynamical evolution of multiple Jupiter-sized planet systems \citep[e.g.,][]{Marzari&Weidenschilling2002,Nagasawa+2008}. 
Fig. 5 of \cite{Nagasawa+2008} showed a similar gap structure in the distribution of the semimajor axes of planets with our results.

More massive stars have more massive planets due to large surface densities (the top panels).
We also show $\langle M_1/M_{\oplus}\rangle$ estimated by Equation (\ref{eq:fit}) in these panels.
The most massive planets are well described by this equation.
By substituting $\Sigma_1\propto M_*$, $C_{M1}\simeq-1/3$ (Table \ref{table:coefficients}), and $C_{M2}$ (Equation (\ref{eq:C_M2})) into Equation (\ref{eq:fit}), we obtain the dependence of the most massive planet on the stellar mass, 
\begin{eqnarray}
	\log{ \biggl< \frac{ M_1}{M_{\oplus}} \biggr>} &\simeq& 
		0.87 \log{\left( \frac{M_*}{M_{\odot}} \right)} \nonumber \\ &&
		+ 1.2 \log{\left( \frac{\Sigma_1(M_{\odot})}{100~\mbox{g~cm}^{-2}} \right)}
		+1.1,%\nonumber \\
		\label{eq:M_Ms_h1}
\end{eqnarray}
where $\Sigma_1(M_{\odot})$ is $\Sigma_1$ at $M_*=M_{\odot}$.

\begin{table}
	\caption{
		Eccentricities normalized by $e_{{\rm esc},M1}$ of massive and scattered planets.
		}\label{table:e_escM1}
	\begin{tabular}{ccc}
		\hline\hline
		Model & Massive planets & Scattered planets \\
		\hline
		Ms1$\Sigma$100b10 & $0.33 \pm 0.29$ & $0.72 \pm 0.61$ \\
		Ms3$\Sigma$30b10 & $0.71 \pm 0.56$ & $1.1 \pm 1.0$ \\
		Ms5$\Sigma$10b10 & $0.83 \pm 0.58$ & $1.5 \pm 1.7$ \\
		\hline
		Ms0$\Sigma$100b10 & $0.37 \pm 0.36$ & $0.84 \pm 0.85$ \\
		Ms2$\Sigma$30b10 & $0.48 \pm 0.38$ & $1.2 \pm 1.1$ \\
		Ms4$\Sigma$10b10 & $0.78 \pm 0.64$ & $1.1 \pm 1.1$ \\
		\hline
	\end{tabular}
\end{table}

The eccentricities and inclinations of the scattered planets are well expressed by $e_{\rm esc}$ of the massive planets.
These $e_{\rm esc}$ can be estimated by substituting Equation (\ref{eq:M_Ms_h1}) into Equation (\ref{eq:e_esc}), 
\begin{eqnarray}
	e_{{\rm esc},M1} &=& 0.20 \left( \frac{M_*}{ M_{\odot} } \right)^{-0.2}
	\left( \frac{\Sigma_1(M_{\odot})}{100~\mbox{g~cm}^{-2}} \right)^{0.4}
	\left( \frac{a}{0.1~\mbox{au}} \right)^{1/2} 
	\nonumber\\ &&\times 
	\left( \frac{\rho}{3 \mbox{~g~cm}^{-3}} \right)^{1/6},
	\label{eq:e_esc_h1}
\end{eqnarray}
which increase as the stellar mass decreases.
Table \ref{table:e_escM1} summarize eccentricities of the massive and scattered planets normalized by $e_{{\rm esc},M1}$. 
In the estimation of $e_{{\rm esc},M1}$, we adopt $a$ values from numerical results.
The massive planets have their eccentricities $\lesssim e_{{\rm esc},M1}$ since these are damped through the giant impacts \citep{Matsumoto+2015, Matsumoto&Kokubo2017}.
The eccentricities of the scattered planets are $\sim e_{{\rm esc},M1}$ with the standard deviation $\sim e_{{\rm esc},M1}$. 
In the Ms0$\Sigma$100b10 model, which is simulations around $1M_{\odot}$ stars, eccentricities of the massive planets are $\langle e \rangle=0.13\pm 0.089$ and those of the scattered planets are $\langle e \rangle=0.20\pm 0.12$.
These eccentricities are higher than those observed by the {\it Kepler} mission \citep[$\sim$~0.01 -- 0.1, e.g.,][]{Fabrycky+2014, Van_Eylen&Albrecht2015,Zheng_JW+2016,Van_Eylen+2019, Mills+2019b}.
This is because we focus on massive super-Earths (Sect. \ref{sect:hd}) and more massive planets cause stronger scattering ($e_{\rm esc}$, Equation (\ref{eq:e_esc})).
Moreover, planets that are located at 0.01~au -- 0.1~au and have high eccentricities and inclinations are affected by the stellar tide.
Their eccentricities and inclinations would be damped by the tidal interaction after the giant impact stage \citep[e.g.,][]{Goldreich&Soter1966, Papaloizou&Terquem2010, Papaloizou+2018}.
We also discuss the relation between final orbital separations and eccentricities from the point of the orbital crossing timescale in Appendix \ref{sect:delta--ecc}.

We consider the radial velocity larger than 1$\mbox{~m~s}^{-1}$ as the observable criterion.
We assume $\sin{i}=1$ in the estimation of the radial velocity amplitudes.
The planets that satisfy this criterion are plotted as the circles and those with smaller radial velocities are triangles in Fig. \ref{Fig:formed_aMei}.
Most of the massive planets and scattered planets located at $a<1$~au in models Ms0$\Sigma$100b10 and Ms1$\Sigma$100b10 are observable.
As the stellar mass decreases, only planets closer to their stars are observable even if the planets have $ \langle M_1\rangle$ mass (the top panels of Fig. \ref{Fig:formed_aMei}).
The radial velocity amplitude of the massive planet ($K_1$) is 
\begin{eqnarray}
	K_1\propto M_1 M_*^{-1/2} a^{-1/2}=M_*^{0.37} a^{-1/2},
\end{eqnarray}
where we use Equation (\ref{eq:M_Ms_h1}) and $M_1\ll M_*$.
The radial velocity amplitude of planets around low-mass stars is small since planets are less massive around low-mass stars.
Unobservable planets with radial velocity amplitudes less than $1\mbox{~m~s}^{-1}$ would exist at $\gtrsim1$~au around M dwarfs.

\subsection{Ejected protoplanets}\label{sect:ejected}

\begin{figure}[htb]
	\includegraphics[width=\columnwidth]{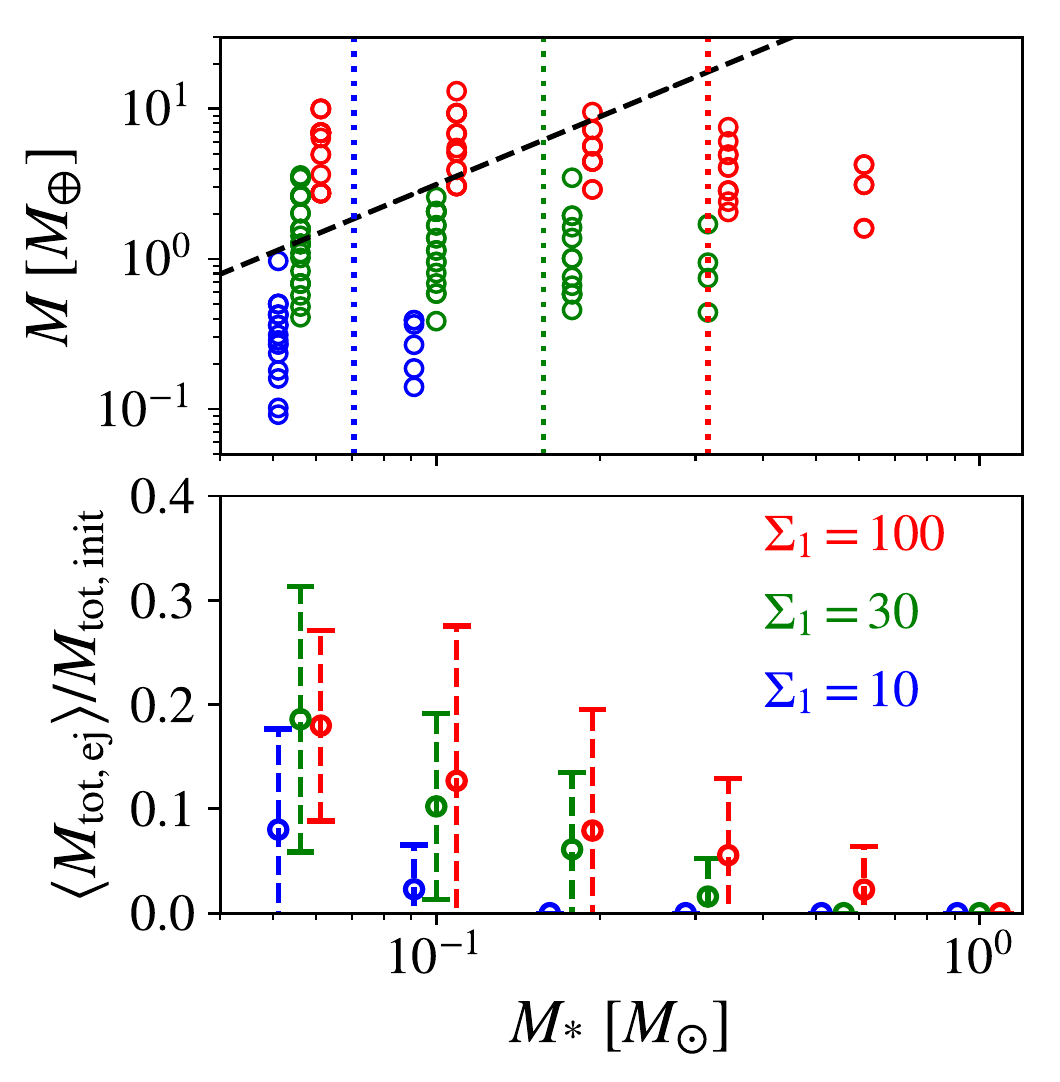}
	\caption{
		Mass of the ejected protoplanets and the ejection rate are shown in the top and bottom panels, respectively.
		The stellar mass in the $\Sigma_1=100~\mbox{g~cm}^{-2}$ and $\Sigma_1=10~\mbox{g~cm}^{-2}$ models is plotted at slightly different $M_*$ to easily compare results.
		The dashed line in the top panel is the ejection mass (Equation (\ref{eq:M_ej})).
		The vertical dotted lines in the top panel are the boundary between the ejection-dominated regime and the collision-dominated regime.
		The error bars in the bottom panel are equal to the standard deviation.
		}
	\label{Fig:Ms_vs_Mej_Mtot}
\end{figure}

Figure \ref{Fig:Ms_vs_Mej_Mtot} shows the mass of each ejected protoplanet and the ejection rate, which is given by the ratio between the total mass of the ejected protoplanets and the total initial mass.
In the $\Sigma_1=100~\mbox{g~cm}^{-2}$ models, the boundary between the ejection and collision-dominated regimes is located around $10^{-1/2}M_{\odot}$.
However, some planets are ejected when $M_*=10^{-1/4}M_{\odot}$ (the Ms1$\Sigma$100b10 model), which is in the collision-dominated regime.
These ejections tend to occur in the outer region ($\gtrsim 1$~au) where $e_{\rm esc}$ is high (Equation (\ref{eq:e_esc})).
The same features are confirmed in the $\Sigma_1=30~\mbox{g~cm}^{-2}$ and $10~\mbox{g~cm}^{-2}$ models.

Masses of the ejected planets are almost in the range of the initial protoplanets (Fig. \ref{Fig:Ms_M1_n}).
In our setting, small-sized protoplanets are initially located close to the central stars.
These protoplanets grow via collisions rather than scatter since their $e_{\rm esc}$ is small (Equation (\ref{eq:e_esc})).
In contrast, the outer large protoplanets have high $e_{\rm esc}$ and strongly scatter each other.
After the inner protoplanets grow via collisions, some outer protoplanets that do not grow and have their initial mass are ejected.

The ejection rate increases as $M_*$ decreases and $\Sigma_1$ increases.
In the ejection-dominated regime, the averaged ejection rates are $\gtrsim 10\%$.
For the surface density with $h_{\rm d}=1$, planets are always ejected in the massive series with 2.2\% -- 8.0\% ejection rates.
In the less-massive series, ejections occur when $10^{-1/2}M_{\odot}$ and $10^{-1}M_{\odot}$, and the ejection rates are 1.6\% and 2.3\%, respectively.
It is plausible that more than a few percentages of the initial mass of protoplanets are ejected from the close-in orbits of low-mass stars in the giant impact stage.

\section{Conclusion}\label{sect:summary}
 
Recent observations have revealed the distribution of close-in super-Earths around M dwarfs.
The mass distribution of close-in super-Earths shows that small planets are orbiting around low-mass stars and $\sim 10$ Earth mass planets are absent around $\sim0.1M_{\odot}$ stars (Fig. \ref{Fig:Ms_M_obs}).
We have investigated the in-situ formation of close-in super-Earths via giant impacts of protoplanets around M dwarfs in a gas-free environment to explain these observed features.
We performed $N$-body simulations with systematically changing the stellar mass and the surface density.
We found that there are two regimes of planetary growth, collision-dominated growth regime and ejection-dominated growth regime.
These regimes are divided by $e_{\rm esc}$, which is calculated as the escape velocity divided by the Kepler velocity.
When $e_{\rm esc}$ are small, planets grow via giant impacts and their growth is in the collision-dominated regime.
In this regime, the largest mass and number of final planets are well expressed by the power-law of the stellar mass and surface density.
When $e_{\rm esc}$ exceeds about 0.4, planets do not grow and some planets are ejected due to several close scattering events.
This regime is the ejection-dominated regime.
In this regime, the largest mass and number of final planets are almost constant.
The boundary of the planet mass between these regimes is $3.0M_{\oplus}$ at 0.1~au around $0.1M_{\odot}$ star (Equation (\ref{eq:M_ej})).
This boundary mass agrees with the mass distribution of the observed close-in super-Earths around low-mass stars (Fig. \ref{Fig:Ms_M_obs}).
This suggests that the initial protoplanets are less massive than this boundary mass and their growth stalls at the boundary mass around low-mass stars.

We performed parameter surveys on the stellar mass ($M_*$) and the surface density of protoplanets ($\Sigma_1$) plane and consider mass and orbital elements of observable planets via the $1\mbox{~m/s}$ radial velocity criterion.
We employed the linear relation, $\Sigma_1\propto M_*$, to reproduce the mass distribution of observed super-Earths.
We considered two series of models and show the distribution of the planets in these series of models.
The mass and orbital distributions of the planets in these two series display similar trends.
The massive planets grow in the range of the initial locations of protoplanets and thus they are located between $\sim 0.1$~au and $\sim1$~au. 
The less massive planets are scattered by the massive planets and they are also located at $\lesssim 0.1$~au or $\gtrsim 1$~au.
The planetary mass is more massive around more massive stars.
In contrast, $e_{\rm esc}$ of the massive planets decreases around more massive stars.
The eccentricities and inclinations of the scattered planets are expressed by $e_{\rm esc}$ of the massive planets.
The eccentricities and inclinations of the massive ones are smaller than those of the scattered ones since their eccentricities and inclinations are damped through giant impacts \citep{Matsumoto+2015,Matsumoto&Kokubo2017}.

We found that ejections of protoplanets occur even in the collision-dominated regime.
Our results suggest that ejection events occur around late M dwarfs and a few protoplanets would be ejected from the close-in orbits around late M dwarfs.

Our results showed that Earth mass planets around low-mass stars are formed via dynamically hot evolution of protoplanets in the giant impact stage.
While we do not consider planetesimals, the giant impact growth of protoplanets would be affected by planetesimals when they damp eccentricities of protoplanets.
The dynamically hot evolution of protoplanets would affect the water delivery of planets by planetesimals since the ice line location is close to the central star around low-mass stars \citep{Ida&Lin2005, Raymond+2007}.
The formation of Earth mass planets around low-mass stars from planetesimals is our future work.

\begin{acknowledgements}

	We thank the anonymous referee for helpful comments.
	This work was achieved using the grant of NAOJ Visiting Joint Research supported by the Research Coordination Committee, National Astronomical Observatory of Japan (NAOJ), National Institutes of Natural Sciences (NINS).
	This research was also supported by MOST in Taiwan through the grant MOST 105-2119-M-001-043-MY3.
	E. K. is supported by JSPS KAKENHI Grant Number 18H05438.
	Numerical simulations were carried out on the PC cluster at the Center for Computational Astrophysics, National Astronomical Observatory of Japan, and at the Academia Sinica Institute for Astronomy and Astrophysics (ASIAA).

\end{acknowledgements}

\bibliographystyle{aa}
\bibliography{bibtex_ym}

\begin{appendix}

\section{Separation--Eccentricity relation}\label{sect:delta--ecc}

\begin{figure}[hbt]
	\resizebox{\hsize}{!}{
		\includegraphics{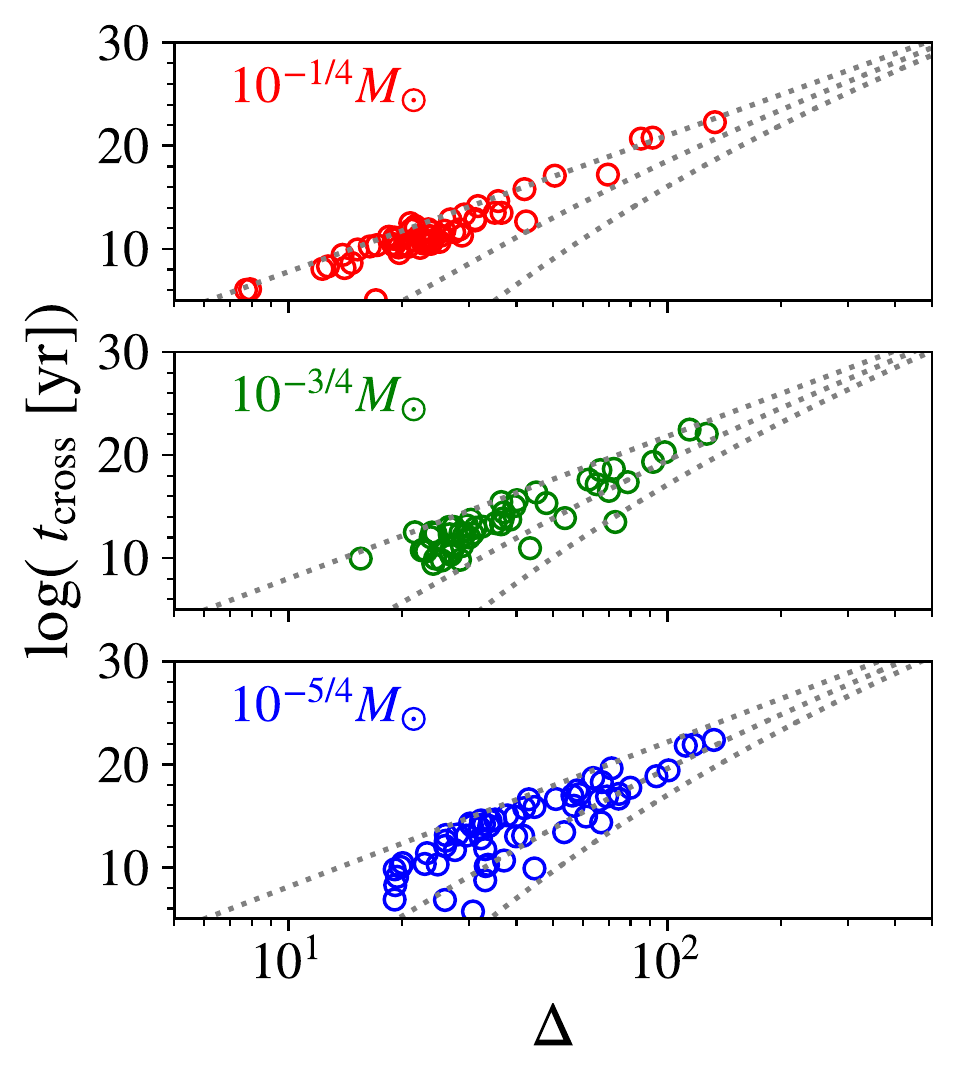}
		\includegraphics{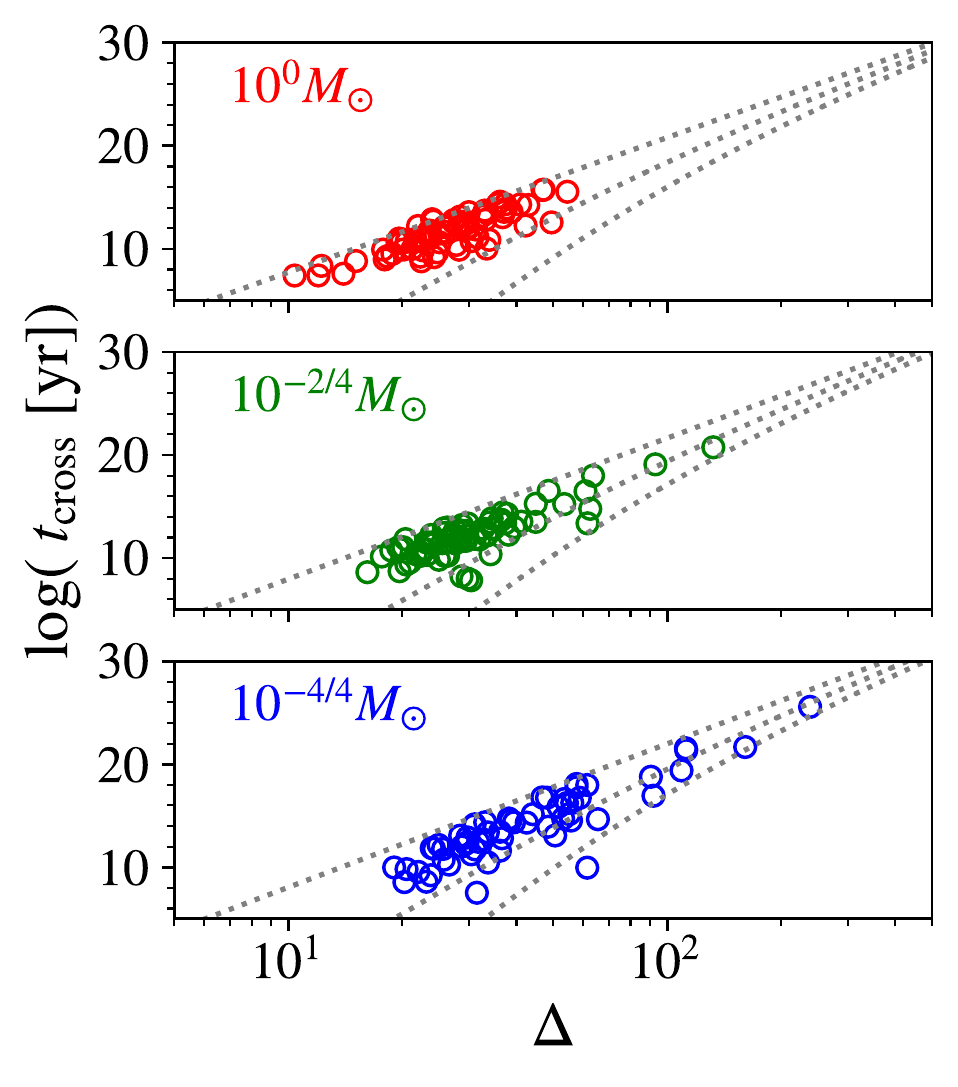}
	}
	\caption{
		Orbital crossing time ($t_{\rm cross}$) of each planet pair is plotted as a function of the orbital separation in mutual Hill radus ($\Delta$).
		The left panels are the results in the massive series (models Ms1$\Sigma$100b10, Ms3$\Sigma$30b10, and Ms5$\Sigma$10b10), and the right ones are those in the less-massive series (models Ms0$\Sigma$100b10, Ms2$\Sigma$30b10, and Ms4$\Sigma$10b10), respectively.
		Three dotted lines are the orbital crossing time with $e=0$ planets (left one), $0.5e_{{\rm esc},M1}$ planets (middle one), and $e_{{\rm esc},M1}$ planets (right one) estimated by Equation (\ref{eq:t_cr}).
		}
	\label{Fig:Delta_tc_col}
\end{figure}

Orbital separations and eccentricities of planets affect their orbital crossing time.
In this section, we investigate their relation.
\cite{Zhou+2007} derived an empirical formula of the orbital crossing time of planets as a function of the orbital separation and eccentricities from $N$-body simulations. 
Their formula was modified to apply to planets whose masses and orbital separations are not identical under the concept of the orbital crossing time of each planet pair by \cite{Ida&Lin2010}.
This estimation of the orbital crossing time of each pair ($t_{\rm cross}$) is
\begin{eqnarray}
	\log{\left( \frac{t_{\rm cross}}{T_{\rm K}} \right)} = A+ B\log{\left( \frac{\Delta}{2.3} \right)}.
	\label{eq:t_cr}
\end{eqnarray}
The coefficients $A$ and $B$ depend on the orbital separation and eccentricities. 
For an adjacent planets pair ($i$, $i+1$), $A$ and $B$ are
\begin{eqnarray}
	A &=& -2+ e_0 -0.27\log{\mu},
	\nonumber \\
	B &=& 18.7+1.1\log{\mu} -(16.8+1.2\log{\mu})e_0, 
\end{eqnarray}
where 
\begin{eqnarray}
	e_0 &=& \frac{e_{i}+e_{i+1}}{2h\Delta},\quad
	\mu = \frac{ (M_i+M_{i+1})/2}{M_*},\quad
	h = \left(\frac{ M_i+M_{i+1} }{3M_*} \right)^{1/3},
	\nonumber \\
	\Delta 
	&=& \frac{ a_{i+1}-a_i }{ R_{\rm H} },
	\quad
	R_{\rm H} = \left( \frac{M_i+M_{i+1}}{3M_*} \right)^{1/3} \left(\frac{a_i+a_{i+1}}{2} \right).
	%\nonumber \\
\end{eqnarray}
We note that this estimation does not include the effect of the number of planets \citep{Chambers+1996,Funk+2010,Matsumoto+2012} and large dispersion of the orbital separations.
\cite{Wu_DH+2019} investigated the orbital crossing time of planets whose $\Delta$ is not uniform and found that their orbital crossing time is equal or longer than the estimation using the minimum separation.
These suggest that the actual crossing times are much longer than the orbital crossing time estimated by Equation (\ref{eq:t_cr}).

We show the orbital crossing time of each adjacent planet pair in planetary systems of the massive and less-massive series (Sect. \ref{sect:hd}) as a function of $\Delta$ in Fig. \ref{Fig:Delta_tc_col}.
We also plot the orbital crossing time with $e=0$, $0.5e_{{\rm esc},M1}$, and $e_{{\rm esc},M1}$ as the dotted lines. 
We find that separations are concentrated in the range from $\Delta=20$ to 30 since the planetary system is stable for much longer than $10^7$~yr when $\gtrsim20R_{\rm H}$.
Almost all orbital crossing times of the planet pairs are between the $e=0$ and $e_{{\rm esc},M1}$ lines.
Specifically, the orbital crossing times in models $10^{-1/4}M_{\odot}$ and $1M_{\odot}$ are located above the $e=0.5e_{{\rm esc},M1}$ line.
As the stellar mass decreases, more planet pairs are distributed between the $e=0.5e_{{\rm esc},M1}$ and $e_{{\rm esc},M1}$ lines.
These distributions arise due to their formation processes.
High $e_{\rm esc}$ of planets around low-mass stars make planets eccentric and their orbital separations wider.

\end{appendix}

\end{document}